\documentclass[
 aip,
 amsmath,amssymb,
 reprint
]{revtex4-1}

\usepackage{graphicx}
\usepackage{dcolumn}
\usepackage{bm}

\usepackage[utf8]{inputenc}
\usepackage[T1]{fontenc}
\usepackage{mathptmx}
\usepackage{etoolbox}

\setcitestyle{super}
\usepackage[font=footnotesize,labelfont=bf,justification=RaggedRight]{caption}
\usepackage{float}
\usepackage{amsmath}
\usepackage{amssymb}
\usepackage{graphicx}
\usepackage{esint}
\usepackage{color}
\usepackage{comment}
\makeatletter    
\graphicspath{ {/} }            
\usepackage{amssymb}
\usepackage{hyperref}

\begin{document}

\preprint{AIP/123-QED}

\title[Microtearding mode study in NSTX using machine learning enhanced reduced model]{Microtearding mode study in NSTX using machine learning enhanced reduced model}

\author{Max T. Curie}
  \affiliation{General Atomics, San Diego, CA, 85608}
  \affiliation{Princeton Plasma Physics Laboratory, Princeton, NJ, 08540}
   \affiliation{Institute for Fusion Studies, University of Texas at Austin, Austin, TX, 	78705}
 \email{xingtian@fusion.gat.com}
 
  \author{Joel Larakers}
 \affiliation{Institute for Fusion Studies, University of Texas at Austin, Austin, TX, 	78705}

 \author{Jason Parisi}
 \affiliation{Princeton Plasma Physics Laboratory, Princeton, NJ, 08540}

 \author{Gary Staebler}
 \affiliation{General Atomics, San Diego, CA, 85608}
 
 \author{Stefano Munaretto}
 \affiliation{Princeton Plasma Physics Laboratory, Princeton, NJ, 08540}
 
 \author{Walter Guttenfelder}
 \affiliation{Princeton Plasma Physics Laboratory, Princeton, NJ, 08540}

 \author{Emily Belli}
 \affiliation{General Atomics, San Diego, CA, 85608}

 \author{David R. Hatch}
 \affiliation{Institute for Fusion Studies, University of Texas at Austin, Austin, TX, 	78705}
 
 \author{Mate Lampert}
 \affiliation{Princeton Plasma Physics Laboratory, Princeton, NJ, 08540}
 
 \author{Galina Avdeeva}
 \affiliation{General Atomics, San Diego, CA, 85608}
 
  \author{Tom Neiser}
 \affiliation{General Atomics, San Diego, CA, 85608}
 
   \author{Sterling Smith}
 \affiliation{General Atomics, San Diego, CA, 85608}

  \author{Ahmed Diallo}
 \affiliation{Princeton Plasma Physics Laboratory, Princeton, NJ, 08540}
 
 \author{Oak Nelson}
   \affiliation{Columbia University, New York, NY, 10027}
   
 \author{Stanley Kaye}
  \affiliation{Princeton Plasma Physics Laboratory, Princeton, NJ, 08540}
  
   \author{Eric Fredrickson}
  \affiliation{Princeton Plasma Physics Laboratory, Princeton, NJ, 08540}
 
\author{Joshua M Manela}
  \affiliation{Michigan Technological University, Houghton, MI, 49931}

\author{Shelly Lei}
  \affiliation{Arizona State University, Tempe, AZ, 85287}

   \author{Michael Halfmoon}
 \affiliation{Institute for Fusion Studies, University of Texas at Austin, Austin, TX, 	78705}
 
 \author{Matthew M Tennery}
 \affiliation{Institute for Fusion Studies, University of Texas at Austin, Austin, TX, 	78705}
 
  \author{Ehab Hassan}
 \affiliation{Oak Ridge National Laboratory, Oak Ridge, TN, 37830}
 \affiliation{Physics, Faculty of Science, Ain Shams University, EG, 11566}
 
  \author{NSTX team}
 \affiliation{Princeton Plasma Physics Laboratory, Princeton, NJ, 08540}

\date{\today}

\begin{abstract}
This article presents a survey of NSTX cases to study the microtearing mode (MTM) stabilities using the newly developed global reduced model for \textbf{S}lab-\textbf{Li}ke \textbf{M}icrotearing modes (SLiM). A trained neutral network version of SLiM enables rapid assessment (0.05s/mode) of MTM with $98\%$ accuracy providing an opportunity for systemic equilibrium reconstructions based on the matching of experimentally observed frequency bands and SLiM prediction across a wide range of parameters. Such a method finds some success in the NSTX discharges, the frequency observed in the experiment matches with what SLiM predicted. Based on the experience with SLiM analysis, a workflow to estimate the potential MTM frequency for a quick assessment based on experimental observation has been established. 
\end{abstract}

\maketitle

\section{Introduction}

National Spherical Torus Experiment (NSTX) is a fusion device based on the spherical Tokamak concept. Studies \cite{doi:10.1063/1.3685698,Guttenfelder_2013,Guttenfelder_2019,Kaye_2021,doi:10.1063/1.4719689,Ren_2013,Ren_2017,Ren_2020, Fenstermacher_2022, doi:10.1063/5.0087403} show the micro-instabilies contributes the degradation of the pedestal through transport. Microtearing mode (MTM) is the electromagnetic micro-instability that is driven by the electron temperature gradient. It contributes a significant amount of electron heat transport in the NSTX pedestal along with electron temperature gradient mode (ETG) \cite{Canik_2013, Maingi_2012}.   

MTM's stability depends on a host of factors \cite{Joel_prl}. Most notably, when the collision frequency is similar to the mode frequency ($\nu_{ei}/\omega\sim 1$), the slab MTM becomes unstable. MTM has mode frequency of the electron diamagnetic frequency at the mode location \cite{Kotschenreuther_2019, Joel_prl}. 
And discharges with lithium-coated plasma-facing components \cite{Canik_2013, MAINGI20151134, MAINGI2017150} provide the collision frequency similar to the diamagnetic frequency $\nu_{ei}/\omega_{*e} \sim 1$, therefore, the slab-like MTM is likely to be unstable. Gyrokinetic simulations found the unstable MTM in the NSTX pedestal and contributed a significant amount of electron heat transport across several discharges. \cite{doi:10.1063/1.4954911,doi:10.1063/5.0011614}

The newly developed \textbf{S}lab \textbf{Li}ke MTM (SLiM) model has successfully demonstrated its application in conventional Tokamaks such as DIII-D and JET \cite{PoP_2021_Curie} on explaining mode skipping, chirping, and calculating the mode frequency and stability. Importantly, due to the high level of sensitivity of location of the rational surfaces safety factor. SLiM has demonstrated its ability to constrain the safety factor at the pedestal which provides a better profile for more computationally costly simulations. 

This article will explore SLiM's capability on NSTX with a more sophisticated profile variation scheme. Along the way, a workflow using SLiM to assist gyrokinetic analysis is provided. And SLiM's limitation in strongly shaped devices such as NSTX is discussed. 

The article has the following structure. 
Chapter \ref{sec:Background} is a brief review of the background of the SLiM, including the past success on conventional Tokamaks. An example from the past article shows its ability to constrain the profile. It is hard to achieve such accuracy with pure experimental observation. 
Chapter \ref{sec:prof_mod} presents the methodology for varying the profile. 
Chapter \ref{sec:SLiM_determine} shows the way SLiM varies the equilibrium and method to pick the equilibrium that has SLiM predicted MTM best matches experimental observed magnetic signals. 
Chapter \ref{sec:application} presents an NSTX case showcase of such a tool for picking the best equilibrium. 
Chapter \ref{sec:workflow} draws the conclusions and presents the workflow inspired by SLiM that can be applied to any reduced models to help guide the high-fidelity simulations. As a supplement, Chapter \ref{sec:NN} will discuss the trained neural network for SLiM in order to speed up the mode identification enough to sample large profile variations, Chapter \ref{sec:def} will shows the definitions of quantities in greater detail. 

\section{Background}\label{sec:Background}

The SLiM model is the linear slab MTM model that uses kinetic theory \cite{Joel_prl}. Such a reduced model solves the dispersion relation defined by Eqs.\ref{eq:ohm} and \ref{eq:sum_q=0}.

\begin{equation}
    \frac{d^2A_{||}}{dx^2}=-\frac{4\pi}{c}\sigma_{||}(\omega,x)E_{||} \label{eq:ohm}
\end{equation}

\begin{equation}
    \left(\frac{c}{v_A}\right)^2(\omega-\omega_{*n})\frac{d^2\phi}{dx^2}=-4\pi k_{||}\sigma_{||}(\omega,x)E_{||}  \label{eq:sum_q=0}
\end{equation}

In the equation above, $A_{||}$ is the magnetic vector potential that is parallel to the magnetic field $B_0$, $\phi$ is the electric potential, $E_{||}$ is the electric field that parallels to $B_0$, $\sigma_{||}(\omega,x)$ is the conductivity\cite{doi:10.1063/5.0006215} parallel to $B_0$, $c$ is the speed of light, $x$ is the distance from the rational surface to the $\omega_{*e}$ peak normalized to gyro-radius, $v_A$ is the Alfven velocity, $k_{||}=\hat{b}\cdot \textbf{k}$, where $\hat{b}$ is the unit vector of the magnetic field $\vec{B_0}$, and Eq.\ref{eq:sum_q=0} is based on quasi-neutrality using kinetic theory.  Eq.\ref{eq:ohm} is derived from Ampere's law and Ohm's law. 

\begin{figure}[h] \centering
        \includegraphics[width=0.45\textwidth]{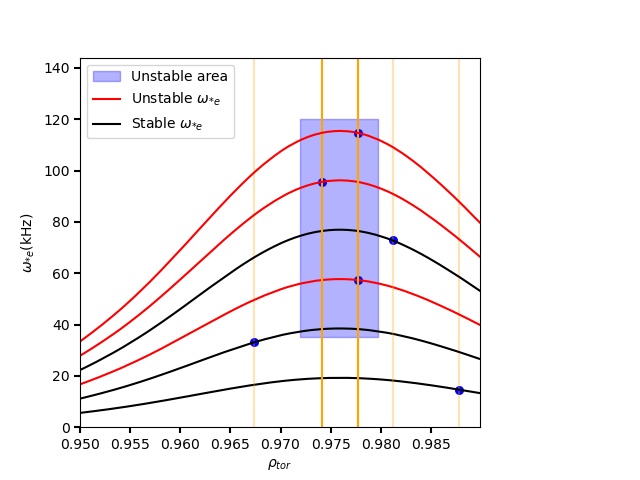}
        \caption{This plot shows the alignment of rational surfaces (orange vertical lines) and $\omega_{*e}$ (black/stable and red/unstable curves), the purple highlighted area is bounded by the frequency observed in experiment and top $4\%$ of the $\omega_{*e}$, the dots represent the intersections of the rational surfaces with the corresponding $\omega_{*e}$ curves. }
        \label{fig:SLiM_intro}
\end{figure}

The consideration of the distance of the rational surface to the $\omega_{*e}$ peak makes SLiM the biggest difference from other MTM reduced models. Such a model has shown success in conventional Tokamaks \cite{PoP_2021_Curie}. Fig. \ref{fig:SLiM_intro} presents a DIII-D discharge studied with SLiM \cite{ehab,Curie_dissertation}. Here is how SLiM works: it takes the equilibrium profiles and calculates electron diamagnetic frequency $\omega_{*e}$ (detailed definitions can be found in Sec.~\ref{sec:appendix}) to find the radial range of interest around $\omega_{*e}$ peak. Takes the rational surfaces that are close to the peak, and calculates its corresponding sets of parameters $\nu,Z_{eff},\eta,\hat{s},\beta,k_y,\mu$ (detailed definitions can be found in Sec.~\ref{sec:appendix}). And then SLiM will calculate the mode stabilities and frequency based on the parameters. 

4 DIII-D discharges and 1 JET discharge have been studied by SLiM \cite{PoP_2021_Curie, Ehab, Halfmoon_MTM}. All discharges have magnetic frequency bands with low mode numbers, and the profiles have low magnetic shears relative to the electron pressure gradient $q(\mu=0)-q(\mu=0.2*x_{*})\sim 1/n$. Where $\mu$ is the distance of the rational surface to the $\omega_{*e}$ peak, and $x_*$ is the spread of $\omega_{*e}$, q is the safety factor, $n$ is the toroidal mode number where the frequency band is found. (detailed definitions can be found in Sec.~\ref{sec:appendix}). Such a low magnetic shear enables rational surfaces to be spatially sparse in the pedestal, which produces discrete frequency bands\cite{PoP_2021_Curie}, other than the board band \cite{RIP_Curie}. SLiM can analyze these cases with a relatively high procession in comparison with global gyrokinetic  simulations. 

\section{Variation of the Equilibrium}\label{sec:prof_mod}

With faster SLiM$\_$NN (more detail in Sec.\ref{sec:NN}), and confidence in SLiM$\_$NN's accuracy, let's now consider how to vary the profile. 

For electron density $n_{e}$, we can carry out modifications with 1 free parameter: $n_{e, scale}$.

\begin{equation}
n_{e}=n_{e 0} \left[1+\left(n_{e, scale}-1\right) weight_{n_{e}}(r)\right]
\end{equation} 

where $n_{e0}$ is the nominal electron density profile, $weight_{n_{e}}(r)=1/2+1/2\cdot tanh\left[(r-r_{top})/width\right]$, $r_{top}$ is the location of the top pedestal, $width$ is the width of pedestal. 

Similarly, for electron temperature $T_e$, we have the following expression.
\begin{equation}
T_{e}=T_{e 0} \left[1+\left(T_{e, scale}-1\right) weight_{_{T_{e}}}(r)\right]
\end{equation} 

\begin{figure}[h] \centering
        \includegraphics[width=0.45\textwidth]{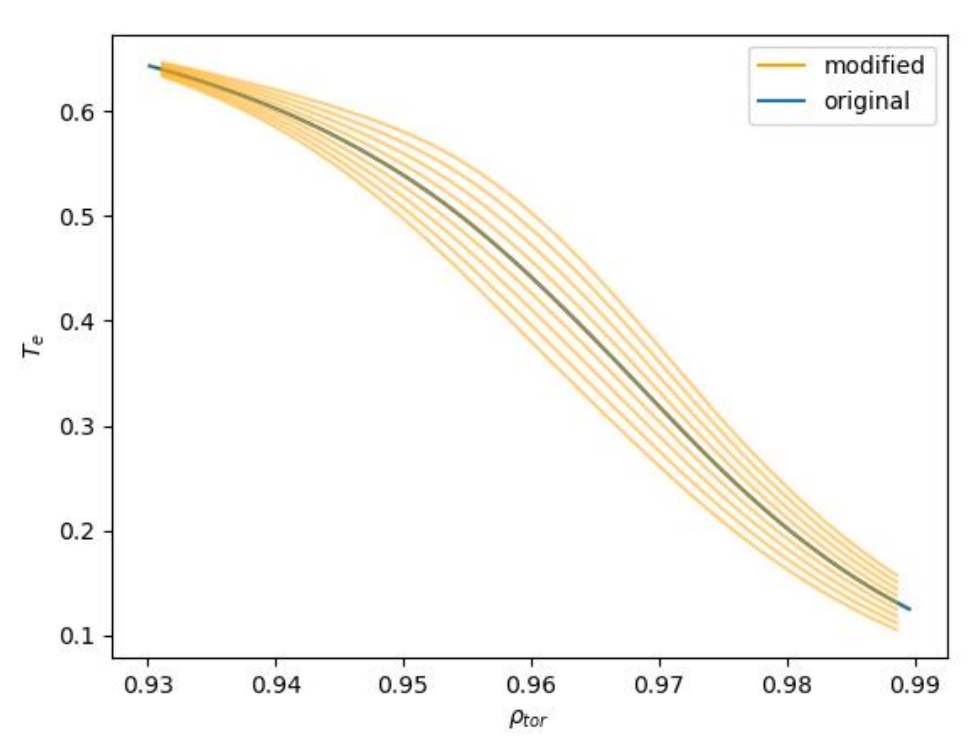}
        \caption{The plot shows the modification of $T_e$ by change $T_{e,scale}$ from $0.8$ to $1.2$ with $0.05$ increment. The blue curve is the nominal profile. The orange lines are the modified profiles, where the bottom curve has $T_{e,scale}=0.8$. }
        \label{fig:Te_scale_mod}
\end{figure}

Where $T_{e0}$ is the nominal electron temperature profile, and the $weight_{T_{e}}(r)=1/2+1/2\cdot tanh\left[(r-r_{top})/width\right]$ is the weight function. 
The weight function is the modified hyperbolic tangent, which goes close to zero at the top pedestal and scrape-off layer (SOL), and near $n_{e,scale}$/$T_{e,scale}$ at mid-pedestal. Thus this modification provides a profile gradient change in the mid-pedestal while not influencing the profile in the core and scrape-off layer (SOL). 
Fig.~\ref{fig:Te_scale_mod} shows the modification of $T_e$ by change $T_{e,scale}$ from $0.8$ to $1.2$ with $0.05$ increment (from bottom to top).

\begin{figure}[h] \centering
        \includegraphics[width=0.45\textwidth]{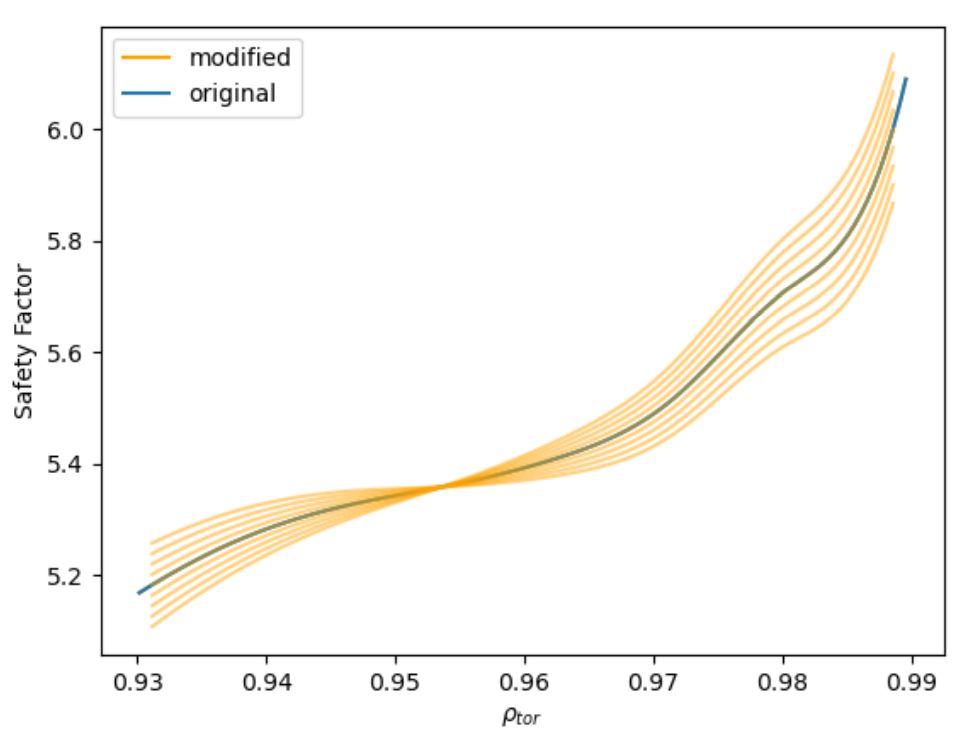}
        \caption{The plot shows the modification of $\hat{s}$ by change $\hat{s}_{scale}$ from $0.8$ to $1.2$ with $0.05$ increment while $q_{scale}=1$, and $q_{shift}=0$. The blue curve is the nominal profile. The orange curves are the modified profile, where the flattest curve around $\rho_{tor}=0.95$ has $\hat{s}_{scale}=0.8$. }
        \label{fig:shat_scale_mod}
\end{figure}

For the safety factor, we can employ similar modifications with 3 free parameters: $\hat{s}_{\text {scale }}$, $q_{scale}$. 

\begin{equation}
q=q_{0} \cdot q_{scale} \cdot\left[1+\left(\hat{s}_{scale}-1\right) weight_{\hat{s}}(r)\right]
\end{equation}

where $weight_{\hat{s} }=-\frac{1}{2} \tanh \left[\frac{r-r_{mid}}{0.1 *  width}\right]$, 0.1 is an arbitrary factor, and $r_{mid}$ is the location of mid-pedestal. To demonstrate the effect of the modification of safety factor with $\hat{s}_{scale}$.  Fig.~\ref{fig:shat_scale_mod} shows the modification of $q$ by change $\hat{s}_{scale}$ from $0.8$ to $1.2$ with $0.05$ increment while $q_{scale}=1$, and $q_{shift}=0$. $\hat{s}_{scale}$ is an important factor to change the spacing between the rational rational surfaces. It is not hard to imagine that the rational surfaces are more densely packed with high $\hat{s}$ while rational surfaces are more sparse given lower $\hat{s}$. Thus lower $\hat{s}$ could help stabilize the unwanted modes by making the rational surfaces further away from the $\omega_{*e}$ peak.

Unless specified, the modification parameter will be kept at nominal values: 
$n_{e,scale}=1$, $T_{e,scale}=1$, $\hat{s}_{\text {scale }}=1$, $q_{scale}=1$. 

\section{Method to determine the best equilibrium}\label{sec:SLiM_determine}

Since a large number of possible equilibria are sampled, manual checking is unrealistic. A metric is constructed to find the optimum equilibrium. A set of magnetic frequency bands can be observed in experiment spectrogram $f_{exp,i}$, where subscript 'exp' stands for the experiment, 'i' stands for the $i_{th}$ frequency band with mode number $n_{tor}=i$. SLiM then can calculate the unstable MTM frequency $f_{SLiM,i}$ based on the equilibrium. We can then construct a metric $\delta_f$ to assess how good the reconstructed equilibrium is based on the frequency matching between the experiment and SLiM calculation. 

\begin{equation}
    \delta_f=\frac{1}{N}\sum_{i=1}^N |f_{exp,i}-f_{SLiM,i}|/f_{exp,i}
\end{equation}

Where $N$ is the total number of frequency bands. We can then choose the reconstructed profile with the smallest $\delta_f$ and use such a profile for further investigations such as large-scale simulations.

\section{Application to discharges}\label{sec:application}

\subsection{NSTX 132588}

Let's use an example to illustrate the method mentioned in the previous section (Sec.~\ref{sec:SLiM_determine})

\begin{figure}[h] \centering
    \includegraphics[width=0.45\textwidth]{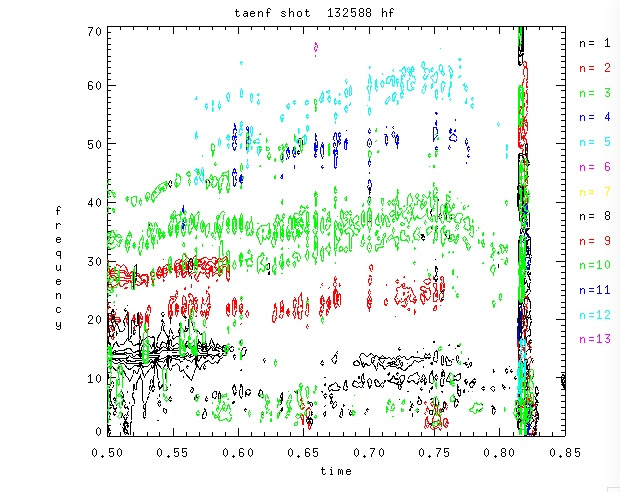}
        \caption{This magnetic spectrogram over from 0.5 seconds to 0.85 seconds with toroidal mode number ranging from 1 to 6}
        \label{fig:mag_132588}
\end{figure}

We can estimate the frequency and mode number from the magnetic spectrogram shown in Fig. \ref{fig:mag_132588}. Such an estimate has been given in the Ch. \ref{ch:132588_freq}. It is worth noting that the only effect magnetic shear will play in SLiM is to change the distance of the rational surfaces. And we are looking for rational surfaces that resonate with $n_{tor}=1$, so scanning $\hat{s}_{scale}$ is not necessary. 

\begin{table}[]
\begin{tabular}{|l|l|}
\hline
mode number & frequency \\ \hline
1           & 14        \\ \hline
2           & 21        \\ \hline
3           & 35        \\ \hline
4           & 50        \\ \hline
5           & 59        \\ \hline
\end{tabular}
\caption{This chart shows the mode number and frequency of each frequency band in the NSTX 132588}
\label{ch:132588_freq}
\end{table}

\begin{figure}[h] \centering
\includegraphics[width=0.45\textwidth]{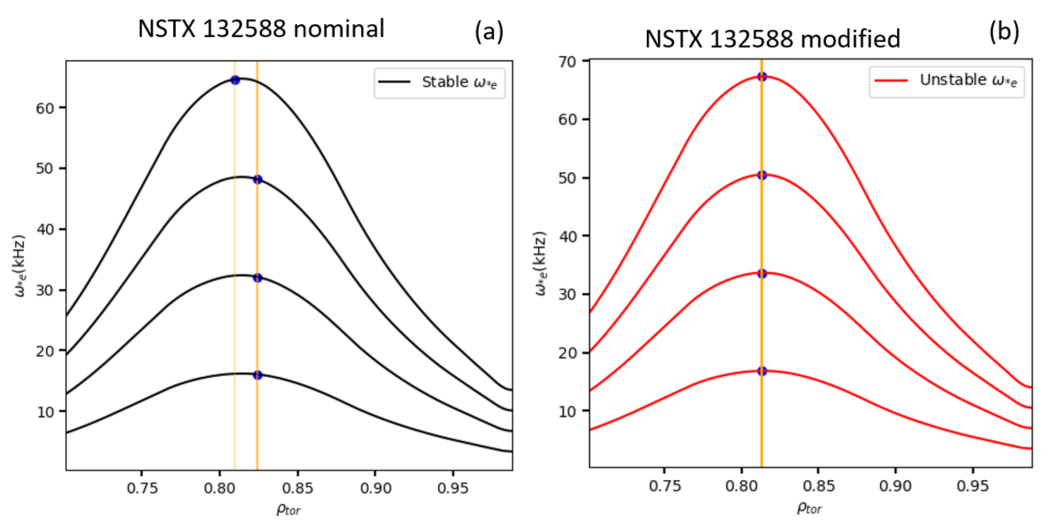}
        \caption{This figure shows the alignment of rational surfaces with $\omega_{*e}$ in plasma frame with nominal profile (figure a) and modified profile (figure b). Where the curves are $\omega_{*e}$ with $n_{tor}=1$ at the bottom with 1 increment. The orange lines are the rational surfaces that intersected with their corresponding $\omega_{*e}$ curves at the blue dots. The red curve means that the mode number contains a potentially unstable MTM, while the black means stable. }
        \label{fig:align_132588}
\end{figure}

The nominal profile does not produce the frequency that agrees with experimental observations shown in  Fig.\ref{fig:align_132588} (a). The frequency is mismatched for the other mode numbers while SLiM does not find unstable MTM since its rational surfaces are too far from the $\omega_{*e}$ peak. 

SLiM takes the following set of variations in Ch. \ref{ch:variation} in order to find the unstable MTM that is the best match with experimentally observed magnetic frequency bands. 

\begin{table}[]
\begin{tabular}{|l|l|l|l|}
\hline
quantity & min & max & increment \\ \hline
$q_{scale}$   & 0.8 & 1.2 & 0.001       \\ \hline
$n_{e,scale}$  & 0.8 & 1.2 & 0.01       \\ \hline
$T_{e,scale}$  & 0.8 & 1.2 & 0.01       \\ \hline
\end{tabular}
\caption{This table describes the variation of profile that SLiM takes to find the best fit with the experiment. }
\label{ch:variation}
\end{table}

\begin{figure}[h] \centering
        \includegraphics[width=0.45\textwidth]{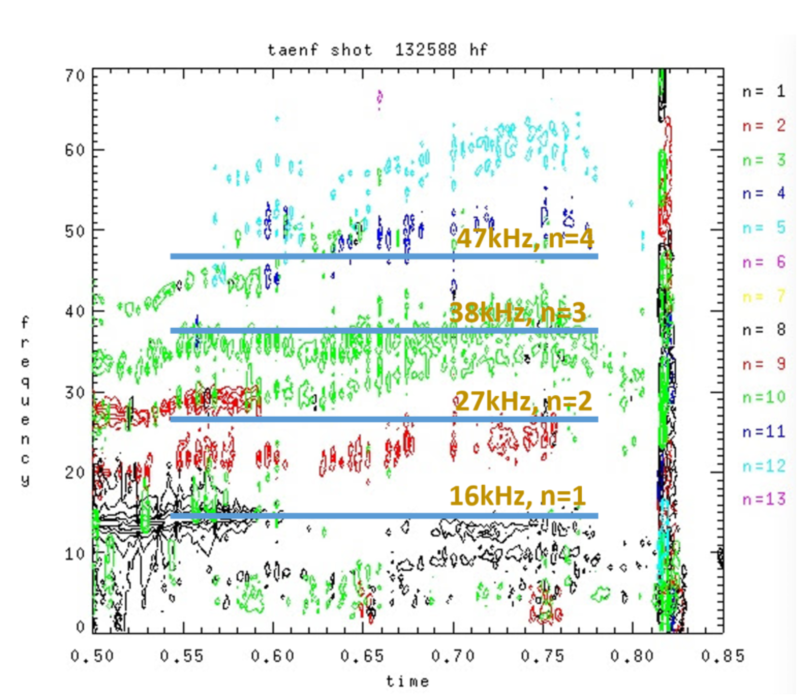}
        \caption{This figure shows the overlay of the frequency (yellow text for frequency and toroidal mode number) of SLiM calculations (blue lines) and experimentally observed magnetic spectrogram.  }
        \label{fig:SLiM_VS_Exp}
\end{figure}

With $q_{scale}=1.04$, $n_{e,scale}=1.12$, $T_{e,scale}=1$, the rational surfaces perfectly align with the $\omega_{*e}$ peak shown in Fig. \ref{fig:align_132588} (b). SLiM find 4 unstable MTM ($n_{tor}=1\sim 4$) that matches with the experiment. Fig. \ref{fig:SLiM_VS_Exp} shows the matching of the SLiM calculation and experimental observation. 

\begin{figure}[h] \centering
        \includegraphics[width=0.45\textwidth]{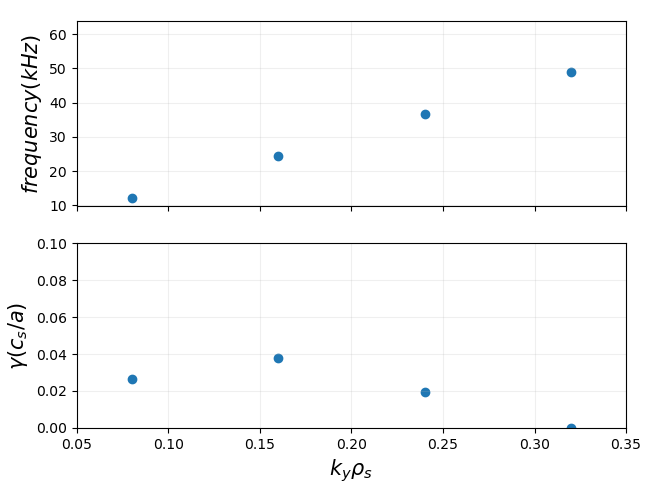}
        \caption{This plot shows the frequency in the lab frame (top plot) and growth rate (bottom plot) of MTM by SLiM calculations. With $k_y\rho_s$ from left to right maps to $n=1 \sim 4$. }
        \label{fig:SLiM_132588}
\end{figure}

The profile is then reproduced with the variation. Fig. \ref{fig:SLiM_132588} shows the frequency and growth rate of the SLiM calculation. 

Additionally, Guttenfelder (2022) \cite{Walter_APS_2022} uses CGYRO nonlinear simulations to show that the MTM can explain the missing electron heat transport in this discharge. 

\subsection{NSTX 129038}\label{sec:case2}
\begin{figure}[h] \centering
    \includegraphics[width=0.45\textwidth]{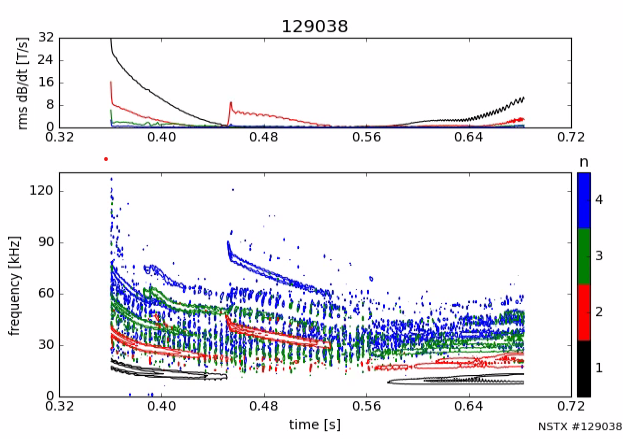}
        \caption{This plot show the magnetic spectrogram, at 500ms, $n_{tor}=2,4$ with $f_2=35kHz$ and $f_4=70kHz$}
        \label{fig:mag_139038}
\end{figure}

NSTX 129038 has a magnetic signal with the mode number $n_{tor}=2,4$ with $f_2=32kHz$ and $f_4=64kHz$ at $t=500ms$, which is shown in Fig. \ref{fig:mag_139038}. 

This case could demonstrate the application of rational surface alignment to constrain $q$ profile in the pedestal. rational surfaces $n_{tor}=4$ exist on top of all $n_{tor}=2$ since it is resonating. Thus, in order to have a profile with $n_{tor}=2,4$ rational surfaces at the $\omega_{*e}$ peak. We need to have $q_{peak}=m/2$, where $m$ is an arbitrary integer. While we do not have rational surfaces with $n_{tor}=1$, we want to avoid $q_{peak}=m$. In order to make the frequency match, and make MTM more unstable, we take $T_{e,scale}=1.2$. Since the $q\sim 10$ around the $\omega_{*e}$ peak. We can take $q=10.5$ using $q_{scale}=1.07$. 
This profile modification makes $\omega_{*e}$ peak align with $n_{tor}=2,4$ rational surfaces. Figure.\ref{fig:align_139038} shows the nominal profile (a) V.S. modified profile (b). The resulting frequency from SLiM matches the experiment as shown in Fig. \ref{fig:align_139038}. 

\begin{figure}[h] \centering
        \includegraphics[width=0.45\textwidth]{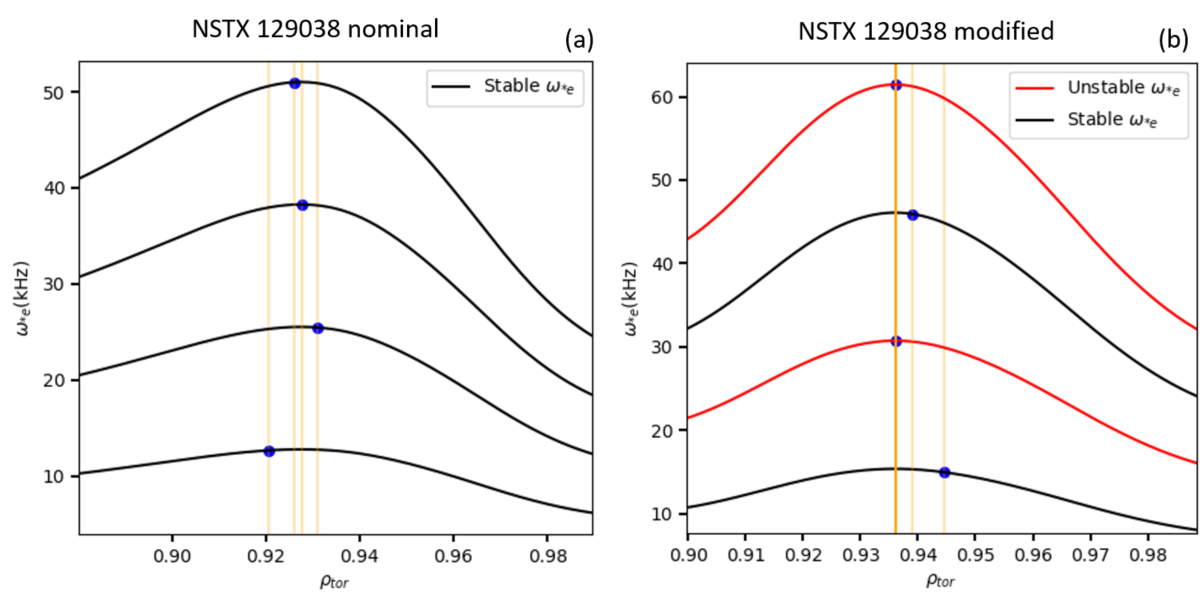}
        \caption{This figure shows the alignment of rational surfaces with $\omega_{*e}$ in plasma frame with nominal profile (figure a) and modified profile (figure b). Where the curves are $\omega_{*e}$ with $n_{tor}=1$ at the bottom with 1 increment. The orange lines are the rational surfaces that intersected with their corresponding $\omega_{*e}$ curves at the blue dots. The red curve means that the mode number contains a potentially unstable MTM, while the black means stable.}
        \label{fig:align_139038}
\end{figure}

\begin{figure}[h] \centering
        \includegraphics[width=0.45\textwidth]{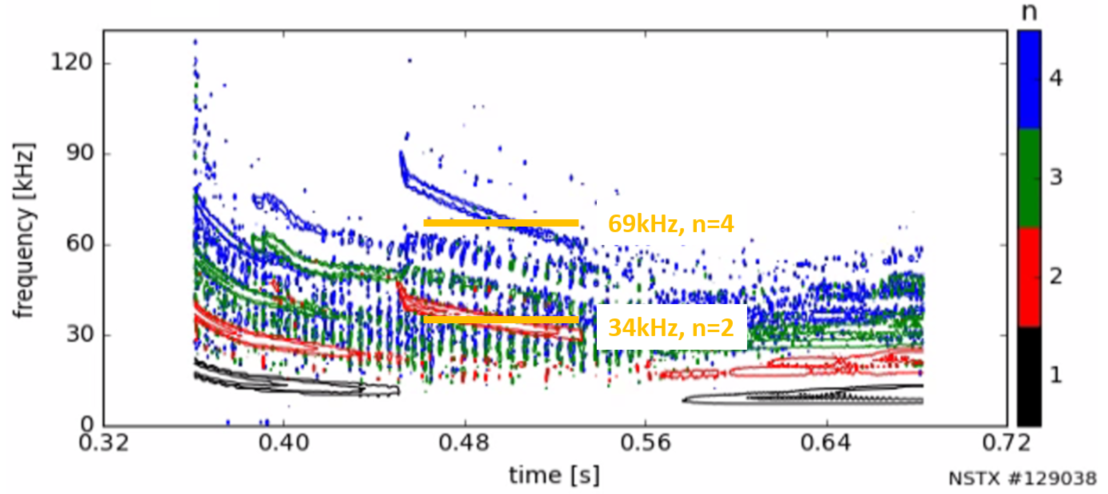}
        \caption{This figure shows the overlay of the frequency (orange text for frequency and toroidal mode number) of SLiM calculations (orange lines) and experimentally observed magnetic spectrogram.  }
        \label{fig:align_139038}
\end{figure}


\section{Discussion and conclusion}\label{sec:workflow}

Notice the NSTX rational surfaces are closer than DIII-D, it is due to higher magnetic shear in NSTX and the less steep electron pressure gradient, thus larger pedestal width. 

Despite the NSTX being strongly shaped, SLiM demonstrated success on several discharges. 
But is important to point out that SLiM works the best at low magnetic shear, with modes number below 10. There are observations of MTM with high toroidal mode number \cite{RIP_Curie,Ehab,doi:10.1063/5.0029996,doi:10.1063/5.0039154}, such higher frequency magnetic frequency bands can be observed in narrow pedestal H-mode NSTX discharges with high collision frequency. 

SLiM neural net enables the large sample size of variation of the profile so that one can get desired profile for further investigations. Two discharges in NSTX are presented to showcase SLiM's capabilities for predicting MTM in spherical Tomkamaks by comparing the results with a magnetic spectrogram using the varied profile produced by SLiM. The shaping effect will be further discussed in a future publication. 

Regardless of the limitations of the SLiM, it can still be a powerful tool to find the right equilibrium for simulations. The workflow can be the following:

\begin{itemize}
    \item Find the mode numbers and mode frequency for the potential MTM in the discharge using a magnetic spectrogram. 
    \item Use the SLiM neural network to find the variation of equilibrium that matches the experimentally observed frequency and mode number. 
    \item Select the desired equilibrium as a reference and reconstruct the equilibrium using self-consistent equilibrium.  
    \item Test the newly reconstructed equilibrium on SLiM    
    \item Conduct further high-fidelity simulations. 
\end{itemize}

The method of profile modification discussed in \ref{ch:variation} is not physically self-consistent. Sample a few profiles that have SLiM prediction matches with experiments to reconstruct new equilibriums will be desirable for the next step of the research. Such a method could potentially aid the equilibrium reconstructions by constraining the profile on the discharges that have the potential slab MTM.  

\section{Appendix}\label{sec:appendix}

\subsection{Training Neutral Network}\label{sec:NN}

\begin{figure}[h] \centering
        \includegraphics[width=0.45\textwidth]{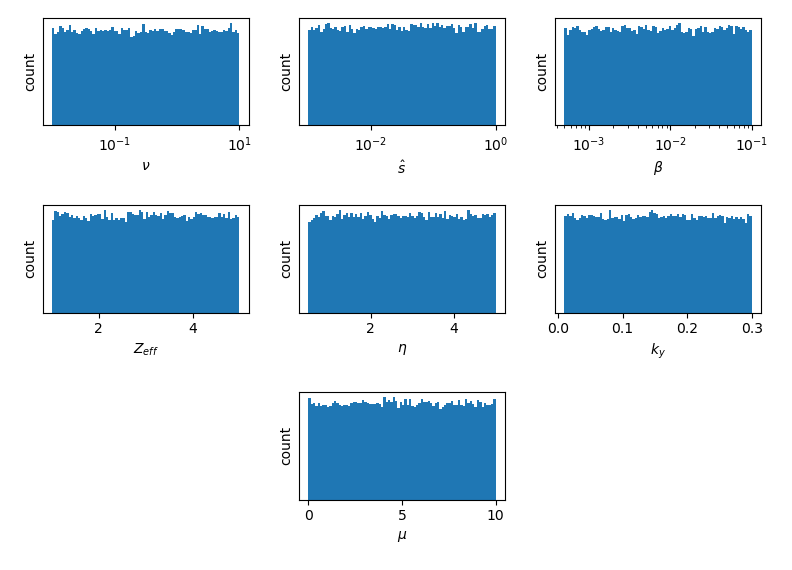}
        \caption{This plot shows the histogram of the training data distribution}
        \label{fig:SLiM_NN_param}
\end{figure}

To sample sufficient variations to the equilibrium to find a match between unstable MTMs calculated by SLiM and experimentally observed magnetic frequency bands, SLiM needed to be sped up. Algorithmic improvements provided a 10-fold increase in speed by using vectorization and simplifying the calculation. However, 15sec/mode is not fast enough for sampling a large set of possible equilibria. Let's do a quick calculation. Assuming 5 modes per equilibrium to be calculated, sampling 2000 equilibria will take 40 hours to finish. The good news is that the improved version of the dispersion calculation is economic enough to run on a high-performance computer over a large parameter range to train a neural network. There are over 3 million dispersion calculations within a normal operating range of DIII-D and NSTX for training. The varying parameters are $\nu,Z_{eff},\eta,\hat{s},\beta,k_y,\mu$ (detailed definitions can be found in Sec.~\ref{sec:appendix}), and keep $x_*=10$. The range of the variable is listed in Table. \ref{ch:slim_NN_param}. To further illustrate the range of variables for the calculation Fig. \ref{fig:SLiM_NN_param}

\begin{table}[H]
\centering
\begin{tabular}{|c|c|c|c|c|c|c|c|}
\hline
Quantity     & $\nu$  & $\hat{s}$ & $\beta$   & $Z_{eff}$   & $\eta$    & $k_y$     & $\mu$     \\ \hline
Minimum      & 0.01 & 0.001 & 0.0005 & 1      & 0.5    & 0.01   & 0      \\ \hline
Maximum      & 10   & 1     & 0.1    & 5      & 5      & 0.3    & 10     \\ \hline
Distribution & log  & log   & log    & linear & linear & linear & linear \\ \hline
\end{tabular}
\caption{The chart shows the range of the variables that SLiM uses for calculation. The first row shows the name of the quantities. The second row and third row show the minimum and maximum values respectively. The fourth row shows the distribution function, and the "log" and "linear" represents even distribution function in log space and linear space}
\label{ch:slim_NN_param}
\end{table}

\begin{figure}[h] \centering
        \includegraphics[width=0.45\textwidth]{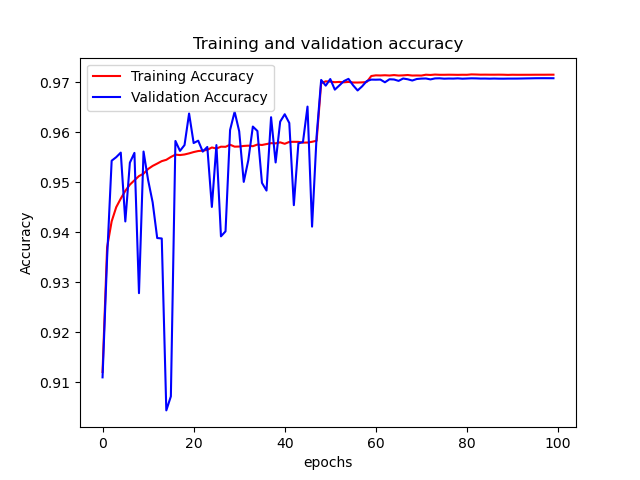}
        \caption{Training and validation accuracy over iterations (epochs) for the training of the neural network of the stability classifier. }
        \label{fig:gamma_acc}
\end{figure}

\begin{figure}[h] \centering
        \includegraphics[width=0.45\textwidth]{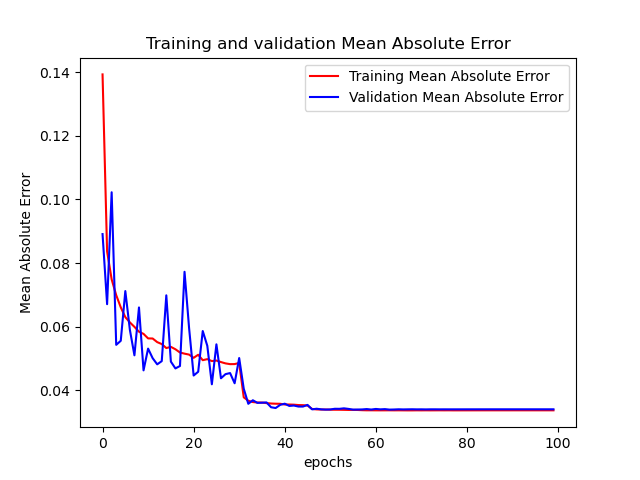}
        \caption{Training and validation mean absolute error over iterations (epochs) for the training of the neural network of the frequency. }
        \label{fig:omega_acc}
\end{figure}

Two neural networks were trained: one is a stability neural network classifier, which categorizes whether the mode has an unstable MTM or not at a given parameter set. The other neural network calculates the frequency of the given unstable MTM. 
Fig.~\ref{fig:gamma_acc} shows the accuracy of the stability prediction of MTM over training iterations (epochs). It shows the validation accuracy of $97.0\%$. And Fig.~\ref{fig:omega_acc} shows the mean average error of MTM frequency prediction over training iterations (epochs), which can be translated into an accuracy of $98.6\%$. 

The trained neural network is accurate, yet 300 times faster than the (already optimized) dispersion calculation. Table.~\ref{ch:MTM_models} shows the updated MTM models including the trained neural network for the SLiM model (SLiM$\_$NN). The trained neural network takes 0.05s/mode to analyze an MTM. Improved speed reduces the sampling of the 2,000 equilibria down to 8 min, which enables a realistic and economic assessment of a large set of variants of the equilibrium. 

\begin{table}[h]
    \centering
    \begin{tabular}{|c|c|c|c|c|c|}
         \hline
        &  &   & & Time  \\
        & Model & Physics & Output &  consumed \\
         &  &   &  &  (sec) \\
         \hline
         & Global linear  & gyrokinetic & moments of & $10^7$  \\
         & simulation  &  &  species &  \\
         \hline
         & SLiM & global slab  & growth rate, & $10^1$\\
         & (dispersion) & model dispersion  & frequency &  \\
         \hline
         & SLiM & neutral network  & stability, & $10^{-2}$\\
         & (trained NN) & on dispersion  & frequency &  \\
         \hline
         & SLiM & alignment of rational & stability & $10^{-4}$\\
         & (alignment) & surfaces to peak  &  & \\
         \hline
\end{tabular}
\caption{The models used in this paper (ranked by the most computationally expensive to the cheapest): Global linear gyrokinetic simulations, SLiM (dispersion calculation), SLiM (trained neutral network from dispersion calculations),  SLiM (alignment of rational surfaces and $\omega_{*e} peak$). }
\label{ch:MTM_models}
\end{table}

The trained neural networks have been benchmarked against the SLiM dispersion calculation. The dispersion relation can be expressed as \\
$\omega(\nu,Z_{eff},\eta,\hat{s},\beta,k_y,\mu/x_*)$ (details can be found in Sec.~\cite{PoP_2021_Curie}).

\begin{figure}[H] \centering
        \includegraphics[width=0.45\textwidth]{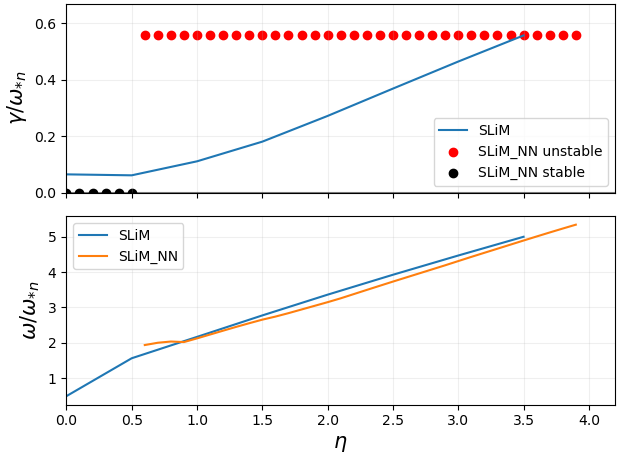}
        \caption{The plot shows the growth rate (top plot) and frequency (bottom plot) with different $\eta$. In the top plot: The blue line shows the growth rate calculated by SLiM. The red dots represent the unstable MTM determined by the neural network version of SLiM: SLiM$\_$NN. The black dots represent the stable MTM determined by the neural network version of SLiM: SLiM$\_$NN, In the bottom plot, the frequency calculated by SLiM (blue line) and SLiM$\_$NN (orange line) are shown}
        \label{fig:eta_scan_NN}
\end{figure}

\begin{figure}[H] \centering
        \includegraphics[width=0.45\textwidth]{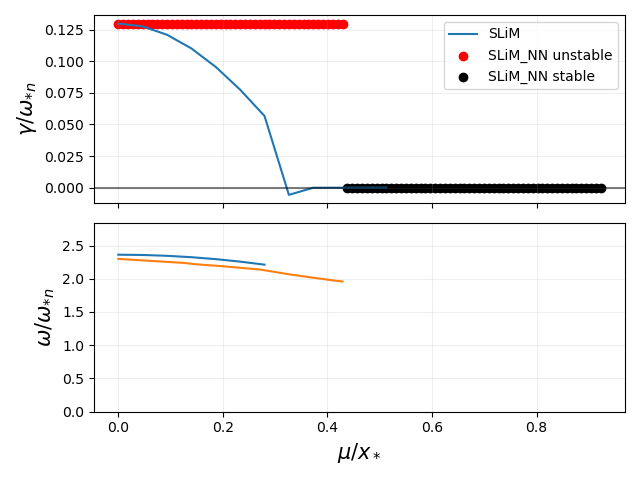}
        \caption{The plot shows the growth rate (top plot) and frequency (bottom plot) with different $\mu$. In the top plot: The blue line shows the growth rate calculated by SLiM. The red dots represent the unstable MTM determined by the neural network version of SLiM: SLiM$\_$NN. The black dots represent the stable MTM determined by the neural network version of SLiM: SLiM$\_$NN, In the bottom plot, the frequency calculated by SLiM (blue line) and SLiM$\_$NN (orange line) are shown}
        \label{fig:mu_scan_NN}
\end{figure}

\begin{figure}[H] \centering
        \includegraphics[width=0.45\textwidth]{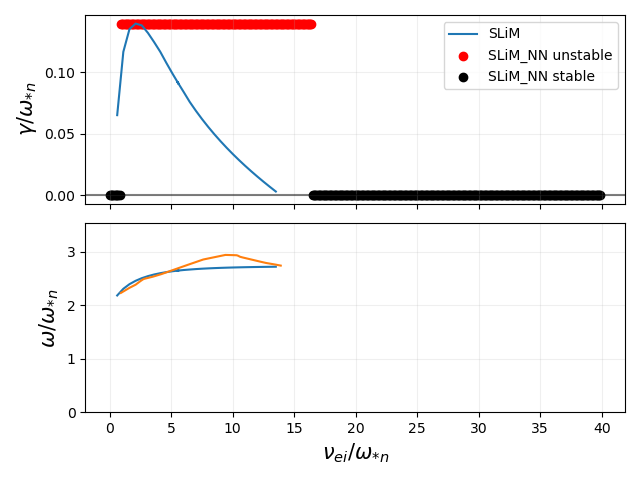}
        \caption{The plot shows the growth rate (top plot) and frequency (bottom plot) with different $\nu$. In the top plot: The blue line shows the growth rate calculated by SLiM. The red dots represent the unstable MTM determined by the neural network version of SLiM: SLiM$\_$NN. The black dots represent the stable MTM determined by the neural network version of SLiM: SLiM$\_$NN, In the bottom plot, the frequency calculated by SLiM (blue line) and SLiM$\_$NN (orange line) are shown}
        \label{fig:nu_scan_NN}
\end{figure}

Benchmarks for an $\eta=\omega_{T}/\omega_{n}$ scan, a $\mu$ (rational surfaces alignment) scan, and a $\nu$ (collision frequency) scan are shown in Fig.~\ref{fig:eta_scan_NN}, Fig.~\ref{fig:mu_scan_NN}, and Fig.~\ref{fig:nu_scan_NN} respectively. The baseline for all three scans is $\nu=1.4$, $Z_{eff}=2.8$, $\eta=1.16$, $\hat{s}=0.006$, $\beta=0.0007$, $k_y\rho_s=0.04$, $\mu/x_*=0$. The plots show the high level of agreement between SLiM$\_$NN and SLiM. Which permits us to proceed to the next step. 

\subsection{Definition of quantities}\label{sec:def}

Here are some definitions of the quantities 

\begin{eqnarray} 
k_y=\sqrt{2}\frac{n_{tor}q\rho_s}{r}\\
\omega_{*e}=\frac{k_y c_s}{\sqrt{2}} \left(\frac{1}{L_{T_e}}+\frac{1}{L_{n_e}}\right)\\
\nu=\nu_{ei}/\omega_{*e,n}\\
Z_{eff}= (n_i+n_zZ^2)/n_e\\
\eta=L_{n_e}/L_{T_e}\\
\hat{s}=L_{n_e}/L_{q}\\
\beta=8\pi n_e k_B T_e/B_0^2
\end{eqnarray} 

Where $\omega_{*e,n}=\frac{n_{tor}q\rho_s c_s}{aL_{n_e}}$, $L_{n_e}=\frac{1}{n_e}\frac{dn_e}{dr}$ is the electron density gradient length scale, 
$L_{T_e}=\frac{1}{T_e}\frac{dT_e}{dr}$ is the electron temperature gradient length scale, 
$L_{q}=\frac{1}{q}\frac{dq}{dr}$ is the safety factor gradient length scale, $c_s=\frac{T_e}{m_i}$ is the speed of sound, $\rho_s=c_s/\omega_g$ is gyro radius, $\omega_g=eB_0/m_ic$ is gyro frequency, $e$ is the electron charge, $q$ is the safety factor, $n_{tor}$ is toroidal mode number, $r$ is the minor radial location, $n_e$ is the electron density, $n_e$ is the ion density, $n_z$ is the impurity density, Z is the charge of the impurity, $k_B$ is the Boltzmann constant, $B_0$ is the magnetic field strength, $m_i$ is the ion mass. 

$\mu$ is the distance from the rational surface to the peak of $\omega_{*e}$, normalized by $\rho_s$, which has been shown in greater detail in \cite{PoP_2021_Curie}.

\section{Acknowledgements}

This material is based upon work supported by the U.S. Department of Energy, Office of Science, Office of Fusion Energy Sciences, using the DIII-D National Fusion Facility, a DOE Office of Science user facility, under Award(s): 
DE-FC02-04ER54698, 
DE-SC0022164. 

This work was supported by U.S. DOE Contract No. DE-FG02-04ER54742 at the Instituted for Fusion Studies (IFS) at the University of Texas at Austin.

This research was supported at Oak Ridge National Laboratory supported by the Office of Science of the U.S. Department of Energy under Contract No. DE-AC05-00OR22725.

This research used resources of the National Energy Research Scientific Computing Center, a DOE Office of Science User Facility. We acknowledge the CINECA award under the ISCRA initiative, for the availability of high performance computing resources and support.

This work was supported by the U.S. Department of Energy under awards DE-SC0022051 and DE-FG02-95ER54309.  This research used resources of the National Energy Research Scientific Computing Center (NERSC), a U.S. Department of Energy Office of Science User Facility located at Lawrence Berkeley National Laboratory, operated under Contract DE-AC02-05CH11231.  This report was prepared as an account of work sponsored by an agency of the United States Government. Neither the United States Government nor any agency thereof, nor any of their employees, makes any warranty, express or implied, or assumes any legal liability or responsibility for the accuracy, completeness, or usefulness of any information, apparatus, product, or process disclosed, or represents that its use would not infringe privately owned rights. Reference herein to any specific commercial product, process, or service by trade name, trademark, manufacturer, or otherwise, does not necessarily constitute or imply its endorsement, recommendation, or favoring by the United States Government or any agency thereof. The views and opinions of authors expressed herein do not necessarily state or reflect those of the United States Government or any agency thereof.


\begin{thebibliography}{1}
\bibitem{parisi_2023}Parisi, J. \& Others Predicting Pedestal Structure With Gyrokinetic Stability and Transport. {\em Physical Review Letters (in Preparation)}. (2023)
\bibitem{doi:10.1063/5.0087403}Hatch, D., Michoski, C., Kuang, D., Chapman-Oplopoiou, B., Curie, M., Halfmoon, M., Hassan, E., Kotschenreuther, M., Mahajan, S., Merlo, G., Pueschel, M., Walker, J. \& Stephens, C. Reduced models for ETG transport in the tokamak pedestal. {\em Physics Of Plasmas}. \textbf{29}, 062501 (2022), 
\bibitem{Fenstermacher_2022}Fenstermacher, M., DIII-D Team:, Abbate, J., Abe, S., Abrams, T., Adams, M., Adamson, B., Aiba, N., Akiyama, T., Aleynikov, P., Allen, E., Allen, S., Anand, H., Anderson, J., Andrew, Y., Andrews, T., Appelt, D., Arbon, R., Ashikawa, N., Ashourvan, A., Aslin, M., Asnis, Y., Austin, M., Ayala, D., Bak, J., Bandyopadhyay, I., Banerjee, S., Barada, K., Bardoczi, L., Barr, J., Bass, E., Battaglia, D., Battey, A., Baumgartner, W., Baylor, L., Beckers, J., Beidler, M., Belli, E., Berkery, J., Bernard, T., Bertelli, N., Beurskens, M., Bielajew, R., Bilgili, S., Biswas, B., Blondel, S., Boedo, J., Bogatu, I., Boivin, R., Bolzonella, T., Bongard, M., Bonnin, X., Bonoli, P., Bonotto, M., Bortolon, A., Bose, S., Bosviel, N., Bouwmans, S., Boyer, M., Boyes, W., Bradley, L., Brambila, R., Brennan, D., Bringuier, S., Brodsky, L., Brookman, M., Brooks, J., Brower, D., Brown, G., Brown, W., Burke, M., Burrell, K., Butler, K., Buttery, R., Bykov, I., Byrne, P., Cacheris, A., Callahan, K., Callen, J., Campbell, G., Candy, J., Canik, J., Cano-Megias, P., Cao, N., Carayannopoulos, L., Carlstrom, T., Carrig, W., Carter, T., Cary, W., Casali, L., Cengher, M., Paz, G., Chaban, R., Chan, V., Chapman, B., Char, I., Chattopadhyay, A., Chen, R., Chen, J., Chen, X., Chen, X., Chen, J., Chen, M., Chen, J., Chen, Z., Choi, M., Choi, W., Choi, G., Chousal, L., Chrobak, C., Chrystal, C., Chung, Y., Churchill, R., Cianciosa, M., Clark, J., Clement, M., Coda, S., Cole, A., Collins, C., Conlin, W., Cooper, A., Cordell, J., Coriton, B., Cote, T., Cothran, J., Creely, A., Crocker, N., Crowe, C., Crowley, B., Crowley, T., Cruz-Zabala, D., Cummings, D., Curie, M., Curreli, D., Molin, A., Dannels, B., Dautt-Silva, A., Davda, K., Tommasi, G., Vries, P., Degrandchamp, G., Degrassie, J., Demers, D., Denk, S., Depasquale, S., Deshazer, E., Diallo, A., Diem, S., Dimits, A., Ding, R., Ding, S., Ding, W., Do, T., Doane, J., Dong, G., Donovan, D., Drake, J., Drews, W., Drobny, J., Du, X., Du, H., Duarte, V., Dudt, D., Dunn, C., Duran, J., Dvorak, A., Effenberg, F., Eidietis, N., Elder, D., Eldon, D., Ellis, R., Elwasif, W., Ennis, D., Erickson, K., Ernst, D., Fasciana, M., Fedorov, D., Feibush, E., Ferraro, N., Ferreira, J., Ferron, J., Fimognari, P., Finkenthal, D., Fitzpatrick, R., Fox, P., Fox, W., Frassinetti, L., Frerichs, H., Frye, H., Fu, Y., Gage, K., Quiroga, J., Gallo, A., Gao, Q., Garcia, A., Munoz, M., Garnier, D., Garofalo, A., Gattuso, A., Geng, D., Gentle, K., Ghosh, D., Giacomelli, L., Gibson, S., Gilson, E., Giroud, C., Glass, F., Glasser, A., Glibert, D., Gohil, P., Gomez, R., Gomez, S., Gong, X., Gonzales, E., Goodman, A., Gorelov, Y., Graber, V., Granetz, R., Gray, T., Green, D., Greenfield, C., Greenwald, M., Grierson, B., Groebner, R., Grosnickle, W., Groth, M., Grunloh, H., Gu, S., Guo, W., Guo, H., Gupta, P., Guterl, J., Guttenfelder, W., Guzman, T., Haar, S., Hager, R., Hahn, S., Halfmoon, M., Hall, T., Hallatschek, K., Halpern, F., Hammett, G., Han, H., Hansen, E., Hansen, C., Hansink, M., Hanson, J., Hanson, M., Hao, G., Harris, A., Harvey, R., Haskey, S., Hassan, E., Hassanein, A., Hatch, D., Hawryluk, R., Hayashi, W., Heidbrink, W., Herfindal, J., Hicok, J., Hill, D., Hinson, E., Holcomb, C., Holland, L., Holland, C., Hollmann, E., Hollocombe, J., Holm, A., Holmes, I., Holtrop, K., Honda, M., Hong, R., Hood, R., Horton, A., Horvath, L., Hosokawa, M., Houshmandyar, S., Howard, N., Howell, E., Hoyt, D., Hu, W., Hu, Y., Hu, Q., Huang, J., Huang, Y., Hughes, J., Human, T., Humphreys, D., Huynh, P., Hyatt, A., Ibanez, C., Ibarra, L., Icasas, R., Ida, K., Igochine, V., In, Y., Inoue, S., Isayama, A., Izacard, O., Izzo, V., Jackson, A., Jacobsen, G., Jaervinen, A., Jalalvand, A., Janhunen, J., Jardin, S., Jarleblad, H., Jeon, Y., Ji, H., Jian, X., Joffrin, E., Johansen, A., Johnson, C., Johnson, T., Jones, C., Joseph, I., Jubas, D., Junge, B., Kalb, W., Kalling, R., Kamath, C., Kang, J., Kaplan, D., Kaptanoglu, A., Kasdorf, S., Kates-Harbeck, J., Kazantzidis, P., Kellman, A., Kellman, D., Kessel, C., Khumthong, K., Kim, E., Kim, H., Kim, J., Kim, S., Kim, J., Kim, H., Kim, K., Kim, C., Kimura, W., King, M., King, J., Kinsey, J., Kirk, A., Kiyan, B., Kleiner, A., Klevarova, V., Knapp, R., Knolker, M., Ko, W., Kobayashi, T., Koch, E., Kochan, M., Koel, B., Koepke, M., Kohn, A., Kolasinski, R., Kolemen, E., Kostadinova, E., Kostuk, M., Kramer, G., Kriete, D., Kripner, L., Kubota, S., Kulchar, J., Kwon, K., Haye, R., Laggner, F., Lan, H., Lantsov, R., Lao, L., Esquisabel, A., Lasnier, C., Lau, C., Leard, B., Lee, J., Lee, R., Lee, M., Lee, M., Lee, Y., Lee, C., Lee, J., Lee, S., Lehnen, M., Leonard, A., Leppink, E., Lesher, M., Lestz, J., Leuer, J., Leuthold, N., Li, X., Li, K., Li, E., Li, G., Li, L., Li, Z., Li, J., Li, Y., Lin, Z., Lin, D., Liu, X., Liu, J., Liu, Y., Liu, T., Liu, Y., Liu, C., Liu, Z., Liu, C., Liu, D., Liu, A., Liu, D., Loarte-Prieto, A., Lodestro, L., Logan, N., Lohr, J., Lombardo, B., Lore, J., Luan, Q., Luce, T., Cortemiglia, T., Luhmann, N., Lunsford, R., Luo, Z., Lvovskiy, A., Lyons, B., Ma, X., Madruga, M., Madsen, B., Maggi, C., Maheshwari, K., Mail, A., Mailloux, J., Maingi, R., Major, M., Makowski, M., Manchanda, R., Marini, C., Marinoni, A., Maris, A., Markovic, T., Marrelli, L., Martin, E., Mateja, J., Matsunaga, G., Maurizio, R., Mauzey, P., Mauzey, D., Mcardle, G., Mcclenaghan, J., Mccollam, K., Mcdevitt, C., Mckay, K., Mckee, G., Mclean, A., Mehta, V., Meier, E., Menard, J., Meneghini, O., Merlo, G., Messer, S., Meyer, W., Michael, C., Michoski, C., Milne, P., Minet, G., Misleh, A., Mitrishkin, Y., Moeller, C., Montes, K., Morales, M., Mordijck, S., Moreau, D., Morosohk, S., Morris, P., Morton, L., Moser, A., Moyer, R., Moynihan, C., Mrazkova, T., Mueller, D., Munaretto, S., Burgos, J., Murphy, C., Murphy, K., Muscatello, C., Myers, C., Nagy, A., Nandipati, G., Navarro, M., Nave, F., Navratil, G., Nazikian, R., Neff, A., Neilson, G., Neiser, T., Neiswanger, W., Nelson, D., Nelson, A., Nespoli, F., Nguyen, R., Nguyen, L., Nguyen, X., Nichols, J., Nocente, M., Nogami, S., Noraky, S., Norausky, N., Nornberg, M., Nygren, R., Odstrcil, T., Ogas, D., Ogorman, T., Ohdachi, S., Ohtani, Y., Okabayashi, M., Okamoto, M., Olavson, L., Olofsson, E., Omullane, M., Oneill, R., Orlov, D., Orvis, W., Osborne, T., Pace, D., Canal, G., Martinez, A., Palacios, L., Pan, C., Pan, Q., Pandit, R., Pandya, M., Pankin, A., Park, Y., Park, J., Park, J., Parker, S., Parks, P., Parsons, M., Patel, B., Pawley, C., Paz-Soldan, C., Peebles, W., Pelton, S., Perillo, R., Petty, C., Peysson, Y., Pierce, D., Pigarov, A., Pigatto, L., Piglowski, D., Pinches, S., Pinsker, R., Piovesan, P., Piper, N., Pironti, A., Pitts, R., Pizzo, J., Plank, U., Podesta, M., Poli, E., Poli, F., Ponce, D., Popovic, Z., Porkolab, M., Porter, G., Powers, C., Powers, S., Prater, R., Pratt, Q., Pusztai, I., Qian, J., Qin, X., Ra, O., Rafiq, T., Raines, T., Raman, R., Rauch, J., Raymond, A., Rea, C., Reich, M., Reiman, A., Reinhold, S., Reinke, M., Reksoatmodjo, R., Ren, Q., Ren, Y., Ren, J., Rensink, M., Renteria, J., Rhodes, T., Rice, J., Roberts, R., Robinson, J., Fernandez, P., Rognlien, T., Rosenthal, A., Rosiello, S., Rost, J., Roveto, J., Rowan, W., Rozenblat, R., Ruane, J., Rudakov, D., Ruiz, J., Rupani, R., Saarelma, S., Sabbagh, S., Sachdev, J., Saenz, J., Saib, S., Salewski, M., Salmi, A., Sammuli, B., Samuell, C., Sandorfi, A., Sang, C., Sarff, J., Sauter, O., Schaubel, K., Schmitz, L., Schmitz, O., Schneider, J., Schroeder, P., Schultz, K., Schuster, E., Schwartz, J., Sciortino, F., Scotti, F., Scoville, J., Seltzman, A., Seol, S., Sfiligoi, I., Shafer, M., Sharapov, S., Shen, H., Sheng, Z., Shepard, T., Shi, S., Shibata, Y., Shin, G., Shiraki, D., Shousha, R., Si, H., Simmerling, P., Sinclair, G., Sinha, J., Sinha, P., Sips, G., Sizyuk, T., Skinner, C., Sladkomedova, A., Slendebroek, T., Slief, J., Smirnov, R., Smith, J., Smith, S., Smith, D., Snipes, J., Snoep, G., Snyder, A., Snyder, P., Solano, E., Solomon, W., Song, J., Sontag, A., Soukhanovskii, V., Spendlove, J., Spong, D., Squire, J., Srinivasan, C., Stacey, W., Staebler, G., Stagner, L., Stange, T., Stangeby, P., Stefan, R., Stemprok, R., Stephan, D., Stillerman, J., Stoltzfus-Dueck, T., Stonecipher, W., Storment, S., Strait, E., Su, D., Sugiyama, L., Sun, Y., Sun, P., Sun, Z., Sun, A., Sundstrom, D., Sung, C., Sungcoco, J., Suttrop, W., Suzuki, Y., Suzuki, T., Svyatkovskiy, A., Swee, C., Sweeney, R., Sweetnam, C., Szepesi, G., Takechi, M., Tala, T., Tanaka, K., Tang, X., Tang, S., Tao, Y., Tao, R., Taussig, D., Taylor, T., Teixeira, K., Teo, K., Theodorsen, A., Thomas, D., Thome, K., Thorman, A., Thornton, A., Ti, A., Tillack, M., Timchenko, N., Tinguely, R., Tompkins, R., Tooker, J., Sousa, A., Trevisan, G., Tripathi, S., Ochoa, A., Truong, D., Tsui, C., Turco, F., Turnbull, A., Umansky, M., Unterberg, E., Vaezi, P., Vail, P., Valdez, J., Valkis, W., Compernolle, B., Galen, J., Kampen, R., Zeeland, M., Verdoolaege, G., Vianello, N., Victor, B., Viezzer, E., Vincena, S., Wade, M., Waelbroeck, F., Wai, J., Wakatsuki, T., Walker, M., Wallace, G., Waltz, R., Wampler, W., Wang, L., Wang, H., Wang, Y., Wang, H., Wang, Z., Wang, H., Wang, Z., Wang, Y., Wang, G., Ward, S., Watkins, M., Watkins, J., Wehner, W., Wei, Y., Weiland, M., Weisberg, D., Welander, A., White, A., White, R., Wiesen, S., Wilcox, R., Wilks, T., Willensdorfer, M., Wilson, H., Wingen, A., Wolde, M., Wolff, M., Woller, K., Wolz, A., Wong, H., Woodruff, S., Wu, M., Wu, Y., Wukitch, S., Wurden, G., Xiao, W., Xie, R., Xing, Z., Xu, X., Xu, C., Xu, G., Yan, Z., Yang, X., Yang, S., Yokoyama, T., Yoneda, R., Yoshida, M., You, K., Younkin, T., Yu, J., Yu, M., Yu, G., Yuan, Q., Zaidenberg, L., Zakharov, L., Zamengo, A., Zamperini, S., Zarnstorff, M., Zeger, E., Zeller, K., Zeng, L., Zerbini, M., Zhang, L., Zhang, X., Zhang, R., Zhang, B., Zhang, J., Zhang, J., Zhao, L., Zhao, B., Zheng, Y., Zheng, L., Zhu, B., Zhu, J., Zhu, Y., Zhu, Y., Zsutty, M. \& Zuin, M. DIII-D research advancing the physics basis for optimizing the tokamak approach to fusion energy. {\em Nuclear Fusion}. \textbf{62}, 042024 (2022,4), https://dx.doi.org/10.1088/1741-4326/ac2ff2
\bibitem{Ehab}Hassan, E., Hatch, D., Halfmoon, M., Curie, M., Kotchenreuther, M., Mahajan, S., Merlo, G., Groebner, R., Nelson, A. \& Diallo, A. Identifying the microtearing modes in the pedestal of DIII-D H-modes using gyrokinetic simulations. {\em Nuclear Fusion}. \textbf{62}, 026008 (2021,12), https://doi.org/10.1088/1741-4326/ac3be5
\bibitem{doi:10.1063/5.0006215}Larakers, J., Hazeltine, R. \& Mahajan, S. A comprehensive conductivity model for drift and micro-tearing modes. {\em Physics Of Plasmas}. \textbf{27}, 062503 (2020), 
\bibitem{Walter_APS_2022}Guttenfelder, W. National Lab Capabilities: PPPL. {\em National Lab Capabilities: PPPL}. (2022), 
\bibitem{Curie_dissertation}Curie, M. Simulations and Reduced Models for Microtearing Modes in the Tokamak Pedestals. {\em Disseration}. (2022), https://doi.org/10.13140/RG.2.2.24468.37769
\bibitem{doi:10.1063/5.0039154}Chen, J., Brower, D., Ding, W., Yan, Z., Curie, M., Kotschenreuther, M., Osborne, T., Strait, E., Hatch, D., Halfmoon, M., Mahajan, S. \& Jian, X. Pedestal magnetic turbulence measurements in ELMy H-mode DIII-D plasmas by Faraday-effect polarimetry. {\em Physics Of Plasmas}. \textbf{28}, 022506 (2021), 
\bibitem{doi:10.1063/5.0029996}Chen, J., Brower, D., Ding, W., Yan, Z., Osborne, T., Strait, E., Curie, M., Hatch, D., Kotschenreuther, M., Jian, X., Halfmoon, M. \& Mahajan, S. Internal measurement of magnetic turbulence in ELMy H-mode tokamak plasmas. {\em Physics Of Plasmas}. \textbf{27}, 120701 (2020), 
\bibitem{Ren_2020}Ren, Y., Wang, W., Guttenfelder, W., Kaye, S., Ruiz-Ruiz, J., Ethier, S., Bell, R., LeBlanc, B., Mazzucato, E., Smith, D., Domier, C. \& Yuh, H. Exploring the regime of validity of global gyrokinetic simulations with spherical tokamak plasmas. {\em Nuclear Fusion}. \textbf{60}, 026005 (2019,12), https://dx.doi.org/10.1088/1741-4326/ab5bf5
\bibitem{Ren_2017}Ren, Y., Belova, E., Gorelenkov, N., Guttenfelder, W., Kaye, S., Mazzucato, E., Peterson, J., Smith, D., Stutman, D., Tritz, K., Wang, W., Yuh, H., Bell, R., Domier, C. \& LeBlanc, B. Recent progress in understanding electron thermal transport in NSTX. {\em Nuclear Fusion}. \textbf{57}, 072002 (2017,3), https://dx.doi.org/10.1088/1741-4326/aa4fba
\bibitem{Halfmoon_MTM}Halfmoon, M., Hatch, D., Kotschenreuther, M., Mahajan, S., Nelson, A., Kolemen, E., Curie, M., Diallo, A., Groebner, R., Hassan, E., Belli, E. \& Candy, J. Gyrokinetic analysis of inter-edge localized mode transport mechanisms in a DIII-D pedestal. {\em Physics Of Plasmas}. \textbf{29}, 112505 (2022), 
\bibitem{RIP_Curie}Curie, M., Hatch, D., Halfmoon, M., Chen, J., Brower, D., Hassan, E., Kotschenreuther, M., Mahajan, S., Groebner, R. \& Team, D. Gyrokinetic simulations compared with magnetic fluctuations diagnosed with a Faraday-effect radial interferometer-polarimeter in the DIII-D pedestal. {\em Nuclear Fusion}. \textbf{62}, 126061 (2022,11), https://dx.doi.org/10.1088/1741-4326/ac9b76
\bibitem{Ren_2013}Ren, Y., Guttenfelder, W., Kaye, S., Mazzucato, E., Bell, R., Diallo, A., Domier, C., LeBlanc, B., Lee, K., Podesta, M., Smith, D. \& Yuh, H. Electron-scale turbulence spectra and plasma thermal transport responding to continuous E × B shear ramp-up in a spherical tokamak. {\em Nuclear Fusion}. \textbf{53}, 083007 (2013,7), https://dx.doi.org/10.1088/0029-5515/53/8/083007
\bibitem{doi:10.1063/1.4719689}Ren, Y., Guttenfelder, W., Kaye, S., Mazzucato, E., Bell, R., Diallo, A., Domier, C., LeBlanc, B., Lee, K., Smith, D. \& Yuh, H. Experimental study of parametric dependence of electron-scale turbulence in a spherical tokamak. {\em Physics Of Plasmas}. \textbf{19}, 056125 (2012), 
\bibitem{Kaye_2021}Kaye, S., Connor, J. \& Roach, C. Thermal confinement and transport in spherical tokamaks: a review. {\em Plasma Physics And Controlled Fusion}. \textbf{63}, 123001 (2021,11), https://dx.doi.org/10.1088/1361-6587/ac2b38
\bibitem{Guttenfelder_2019}Guttenfelder, W., Kaye, S., Kriete, D., Bell, R., Diallo, A., LeBlanc, B., McKee, G., Podesta, M., Sabbagh, S. \& Smith, D. Initial transport and turbulence analysis and gyrokinetic simulation validation in NSTX-U L-mode plasmas. {\em Nuclear Fusion}. \textbf{59}, 056027 (2019,4), https://dx.doi.org/10.1088/1741-4326/ab0b2c
\bibitem{Guttenfelder_2013}Guttenfelder, W., Peterson, J., Candy, J., Kaye, S., Ren, Y., Bell, R., Hammett, G., LeBlanc, B., Mikkelsen, D., Nevins, W. \& Yuh, H. Progress in simulating turbulent electron thermal transport in NSTX. {\em Nuclear Fusion}. \textbf{53}, 093022 (2013,8), https://dx.doi.org/10.1088/0029-5515/53/9/093022
\bibitem{doi:10.1063/1.3685698}Guttenfelder, W., Candy, J., Kaye, S., Nevins, W., Bell, R., Hammett, G., LeBlanc, B. \& Yuh, H. Scaling of linear microtearing stability for a high collisionality National Spherical Torus Experiment discharge. {\em Physics Of Plasmas}. \textbf{19}, 022506 (2012), 
\bibitem{PoP_2021_Curie}Curie, M., Larakers, J., Hatch, D., Nelson, A., Diallo, A., Hassan, E., Guttenfelder, W., Halfmoon, M., Kotschenreuther, M., Hazeltine, R., Mahajan, S., Groebner, R., Chen, J., Thun, C., Frassinetti, L., Saarelma, S., Giroud, C. \& Tennery, M. A survey of pedestal magnetic fluctuations using gyrokinetics and a global reduced model for microtearing stability. {\em Physics Of Plasmas}. \textbf{29}, 042503 (2022), 
\bibitem{Canik_2013}Canik, J., Guttenfelder, W., Maingi, R., Osborne, T., Kubota, S., Ren, Y., Bell, R., Kugel, H., LeBlanc, B. \& Souhkanovskii, V. Edge microstability of NSTX plasmas without and with lithium-coated plasma-facing components. {\em Nuclear Fusion}. \textbf{53}, 113016 (2013,9), https://doi.org/10.1088/0029-5515/53/11/113016
\bibitem{Kotschenreuther_2019}Kotschenreuther, M., Liu, X., Hatch, D., Mahajan, S., Zheng, L., Diallo, A., Groebner, R., Hillesheim, J., Maggi, C., Giroud, C., Koechl, F., Parail, V., Saarelma, S., Solano, E., Chankin, A. \& And Gyrokinetic analysis and simulation of pedestals to identify the culprits for energy losses using `fingerprints'. {\em Nuclear Fusion}. \textbf{59}, 096001 (2019,7), https://doi.org/10.1088/1741-4326/ab1fa2
\bibitem{Joel_prl}Larakers, J., Curie, M., Hatch, D., Hazeltine, R. \& Mahajan, S. Global Theory of Microtearing Modes in the Tokamak Pedestal. {\em Phys. Rev. Lett.}. \textbf{126}, 225001 (2021,6), https://link.aps.org/doi/10.1103/PhysRevLett.126.225001
\bibitem{doi:10.1063/1.3592519}Canik, J., Maingi, R., Kubota, S., Ren, Y., Bell, R., Callen, J., Guttenfelder, W., Kugel, H., LeBlanc, B., Osborne, T. \& Soukhanovskii, V. Edge transport and turbulence reduction with lithium coated plasma facing components in the National Spherical Torus Experiment. {\em Physics Of Plasmas}. \textbf{18}, 056118 (2011), 
\bibitem{Maingi_2012}Maingi, R., Boyle, D., Canik, J., Kaye, S., Skinner, C., Allain, J., Bell, M., Bell, R., Gerhardt, S., Gray, T., Jaworski, M., Kaita, R., Kugel, H., LeBlanc, B., Manickam, J., Mansfield, D., Menard, J., Osborne, T., Raman, R., Roquemore, A., Sabbagh, S., Snyder, P. \& Soukhanovskii, V. The effect of progressively increasing lithium coatings on plasma discharge characteristics, transport, edge profiles and ELM stability in the National Spherical Torus Experiment. {\em Nuclear Fusion}. \textbf{52}, 083001 (2012,6), https://doi.org/10.1088/0029-5515/52/8/083001
\bibitem{doi:10.1063/1.4954911}Coury, M., Guttenfelder, W., Mikkelsen, D., Canik, J., Canal, G., Diallo, A., Kaye, S., Kramer, G. \& Maingi, R. Linear gyrokinetic simulations of microinstabilities within the pedestal region of H-mode NSTX discharges in a highly shaped geometry. {\em Physics Of Plasmas}. \textbf{23}, 062520 (2016), 
\bibitem{MAINGI20151134}Maingi, R., Osborne, T., Bell, M., Bell, R., Boyle, D., Canik, J., Diallo, A., Kaita, R., Kaye, S., Kugel, H., LeBlanc, B., Sabbagh, S., Skinner, C. \& Soukhanovskii, V. Dependence of recycling and edge profiles on lithium evaporation in high triangularity, high performance NSTX H-mode discharges. {\em Journal Of Nuclear Materials}. \textbf{463} pp. 1134-1137 (2015), https://www.sciencedirect.com/science/article/pii/S0022311514007685, PLASMA-SURFACE INTERACTIONS 21
\bibitem{MAINGI2017150}Maingi, R., Canik, J., Bell, R., Boyle, D., Diallo, A., Kaita, R., Kaye, S., LeBlanc, B., Sabbagh, S., Scotti, F. \& Soukhanovskii, V. Effect of progressively increasing lithium conditioning on edge transport and stability in high triangularity NSTX H-modes. {\em Fusion Engineering And Design}. \textbf{117} pp. 150-156 (2017), https://www.sciencedirect.com/science/article/pii/S0920379616304550
\bibitem{doi:10.1063/5.0011614}Battaglia, D., Guttenfelder, W., Bell, R., Diallo, A., Ferraro, N., Fredrickson, E., Gerhardt, S., Kaye, S., Maingi, R. \& Smith, D. Enhanced pedestal H-mode at low edge ion collisionality on NSTX. {\em Physics Of Plasmas}. \textbf{27}, 072511 (2020), 

\end{thebibliography}
\end{document}